\documentclass[twocolumn,superscriptaddress,floatfix,preprintnumbers,amssymb,amsmath]{revtex4-2}
\usepackage[dvips]{graphicx}% Include figure files
\usepackage{dcolumn}% Align table columns on decimal point
\usepackage{bm}% bold math

\usepackage[dvips]{graphicx}% Include  files
\usepackage{dcolumn}% Align table columns on decimal point
\usepackage{color}
\usepackage{bm}% bold math

\usepackage{epstopdf}

\begin{document}
\title{Influence of device non-uniformities on the accuracy of Coulomb blockade thermometry}

\author{Jukka P. Pekola}
\affiliation{Pico group, QTF Centre of Excellence, Department of Applied Physics, Aalto University School of Science, P.O. Box 13500, 00076 Aalto, Finland}

%\affiliation{Moscow Institute of Physics and Technology, 141700 Dolgoprudny, Russia}
\author{Eemil Praks}
\affiliation{Pico group, QTF Centre of Excellence, Department of Applied Physics, Aalto University School of Science, P.O. Box 13500, 00076 Aalto, Finland}
\author{Nikolai Yurttag\"ul}
\affiliation{VTT Technical Research Centre of Finland Ltd, P.O. Box 1000, FI-02044 VTT Espoo, Finland}
\author{Bayan Karimi}
\affiliation{Pico group, QTF Centre of Excellence, Department of Applied Physics, Aalto University School of Science, P.O. Box 13500, 00076 Aalto, Finland}
\affiliation{QTF Centre of Excellence, Department of Physics, Faculty of Science, University of Helsinki, Finland}

\begin{abstract}
We investigate temperature errors of Coulomb blockade thermometer (CBT) arising from inevitable non-uniformities in tunnel junction arrays. The errors are proportional to the junction resistance variance in the universal operation regime and this result holds approximately also beyond this originally studied high temperature range. We present both analytical and numerical results, and discuss briefly their implications on achievable uniformity based on state-of-the-art fabrication of sensors. 
\end{abstract}
\date{\today}
\maketitle

\section{Introduction}
Coulomb blockade thermometer (CBT) \cite{cbt94} has proven to provide calibration-free thermometry over a wide range from sub-mK up to 70 K \cite{basel2021,delft2020,meschke,Bradley2016,Bradley2017,Palma2017} temperatures $T$, i.e., over five decades. Its operation is based on bias voltage $V$ dependent conductance $G$ of an array of tunnel junctions under the competition between single-electron charging effects (energy scale $E_{\rm C}$) and thermal energy $k_{\rm B}T$. The ideal operation range is when $E_{\rm C} \ll k_{\rm B}T$; in this linear regime a universal relation \cite{cbt94}
\begin{equation} \label{cbt1}
V_{1/2}\simeq 5.439 N k_{\rm B}T/e
\end{equation}
holds, where $V_{1/2}$ is the full width at half-minimum of the conductance dip around zero bias voltage and $N$ is the number of junctions in series in the array. One may design the thermometer sensor such that the above relation between $E_{\rm C}$ and $k_{\rm B}T$ is favorable in the temperature range of interest by engineering the parameters in the array accordingly since $E_{\rm C}$, being the charging energy, is inversely proportional to the effective capacitance which is controlled by physical dimensions. Equation~\eqref{cbt1} is a basis for primary thermometry (calibration-free). When the condition $E_{\rm C} \ll k_{\rm B}T$ is compromised, the CBT still yields calibration free thermometry down to much lower temperatures, but with a modified relation between $V_{1/2}$ and $T$ to be discussed below. 
\begin{figure*}
	\centering
	\includegraphics [width=\textwidth] {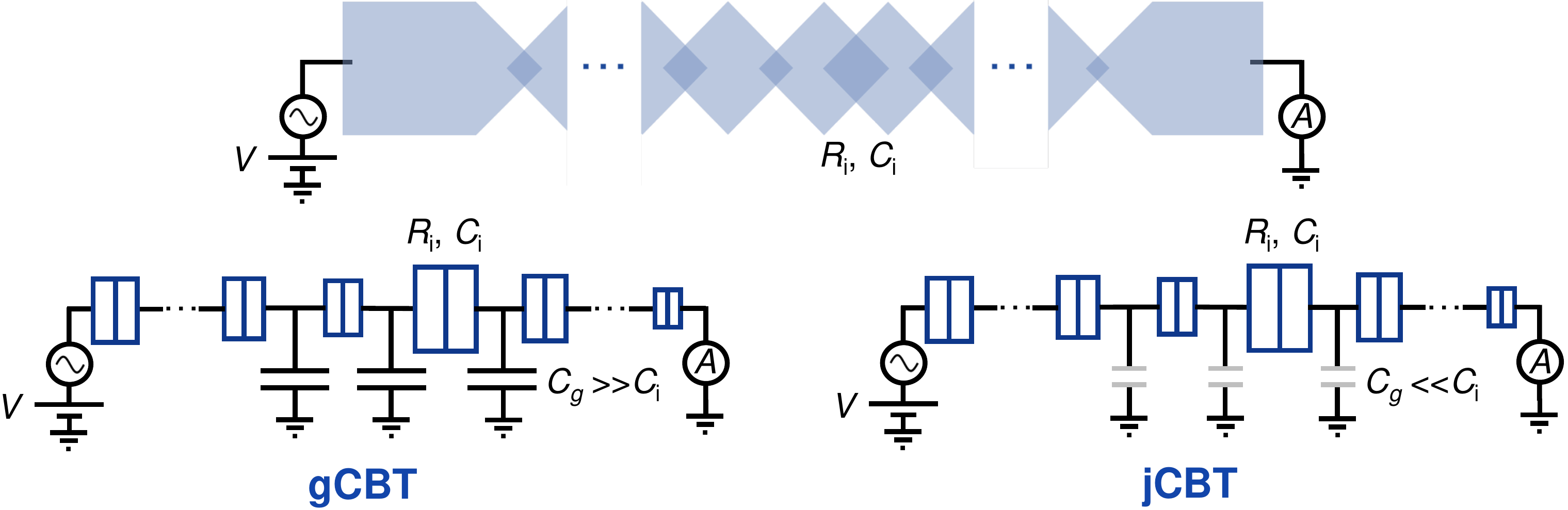}
	\caption{Conceptual illustration of a non-uniform CBT array in the top panel. Two types of CBT are presented in the lower panels where the ground capacitance $C_g$ of each island is either dominating (gCBT on the left) or negligible (jCBT on the right) with respect to junction capacitances $C_i$.
		\label{fig1}}
\end{figure*}

Equation \eqref{cbt1} and its low temperature versions \cite{jltp1997} are strictly valid only for a fully uniform array, where all junction resistances $R_{{\rm T},i}$ are equal through the sensor. Figure~\ref{fig1} (top) shows schematically an array of junctions whose sizes vary along the chain. A way of circumventing the issue of non-uniformity errors is to measure a single junction embedded in a four probe configuration within junction arrays was analyzed and experimentally demonstrated in \cite{prl2008}, and this configuration provides a partial solution to the problem. In fact it yields a fully error-free thermometer in this respect, but the other side of the coin is that the signal in terms of voltage $V_{1/2}$ is small for such a set-up since $N=1$ in Eq. \eqref{cbt1}. Fortunately the effect of non-uniformity is weak in ordinary CBT arrays. However it poses a source of fundamental error, which is the topic of this article. Some aspects of the problem have been addressed previously in Refs. \cite{jltp1997,Hirvi1995}.

In this work we present a comprehensive picture of the non-uniformity errors in Coulomb blockade thermometry. We discuss two different device classes,  (i) the ones where junction resistances vary along the array but capacitance variances are negligible, recently coined gCBT, and (ii) those where both resistances and capacitances vary but such that their product remains constant for each element composed of a junction and island between the junctions, jCBT. These two types of CBT are shown schematically in Fig.~\ref{fig1}. Pure capacitance variations with uniform resistances do not lead to temperature errors in the universal regime, but only to renormalization of charging energy. Devices of type (i) are the ones commonly employed in very low temperature thermometry \cite{basel2021,delft2020,lancaster}, where self-capacitance of the islands between junctions is intentionally increased to bring $E_{\rm C}$ down in order to satisfy the conditions discussed above down to low temperatures. On the other hand class (ii) refers to sensors in higher temperature regimes where junction capacitances dominate over self-capacitances of the islands. We take the product of resistance and capacitance to be constant, since the former one is inversely proportional to the overlap area of the junction, whereas the latter one is proportional to it. Naturally numerical analysis is possible also for arrays that fall being intermediate between these two classes, and their properties can be addressed at least numerically. The general observation is that the non-uniformity errors are proportional to the variance of the parameters in all the situations that we consider. Secondly, we find that the analytical results in the universal regime $E_{\rm C} \ll k_{\rm B}T$ for the error in temperature reading stay approximately valid, also far beyond this domain, based on our numerical results and an analytic calculation to be presented below. Note that all these results apply also for a CBT sensor consisting of several parallel arrays, as commonly used in the experiments. This is because the errors depend only on the variance of the parameters of the sensor. Another point to note here is that we refer to the low and high temperature regimes meaning either to absolute temperature or alternatively to that with respect to $E_{\rm C}$, depending on the context. 
%At the end of the paper we also present experimental tests of non-uniformity in order to assess the practical errors in typical sensors.

\section{Linear regime} \label{linear}
We first consider a CBT array of junctions in the universal regime $E_{\rm C}\ll k_{\rm B}T$. The conductance $G_i$ of junction $i$ normalized by its inverse tunnel resistance $G_{{\rm T},i}=1/R_{{\rm T},i}$ can be written as \cite{Hirvi1995,prl2008}
\begin{equation} \label{e1}
G_i/G_{{\rm T},i}=1-\frac{\delta_i}{k_{\rm B}T}g(v_i),
\end{equation}
where $v_i=eV_i/k_{\rm B}T$ is the voltage $V_i$ across junction $i$ in normalized form, $\delta_i$ arises from the capacitance matrix of the surrounding circuit (not dependent on resistances of the junctions) and $g(x)=e^x[e^x(x-2)+x+2]/(e^x-1)^3$. Based on current conservation through the array and noting that the bias voltage across the whole array is $V=\sum_i V_i$, we find the normalized conductance of the array, $G(V)$ up to linear order in $(k_{\rm B}T)^{-1}$ as
\begin{equation} \label{univ}
G(V)/G_{{\rm T}}=1-\sum_i \frac{R_{{\rm T},i}}{R_\Sigma}\frac{\delta_i}{k_{\rm B}T}g(\frac{R_{{\rm T},i}}{R_\Sigma}\frac{eV}{k_{\rm B}T}).
\end{equation}
Here $R_\Sigma = \sum_i R_{{\rm T},i}\equiv G_{{\rm T}}^{-1}$ is the total resistance of the array. According to this expression, only the resistance non-uniformity affects the absolute temperature reading of the CBT determined by the half-width of the conductance dip, whereas capacitance non-uniformity alone is ineffective.

We expand Eq. \eqref{e1} up to second order in the relative deviations of junction resistances, $\rho_i = R_{{\rm T},i}/R_{\rm ave}-1$, where $R_{\rm ave}=R_\Sigma/N$. We consider two relevant cases (i, gCBT) and (ii, jCBT) described above. Below we normalize all the voltages such that $v=eV/(Nk_{\rm B}T)$ and $v_{1/2}=eV_{1/2}/(Nk_{\rm B}T)$, and denote by $\langle \rho^2\rangle$ the variance of $\rho_i$.

\begin{figure}
	\centering
	\includegraphics [width=\columnwidth] {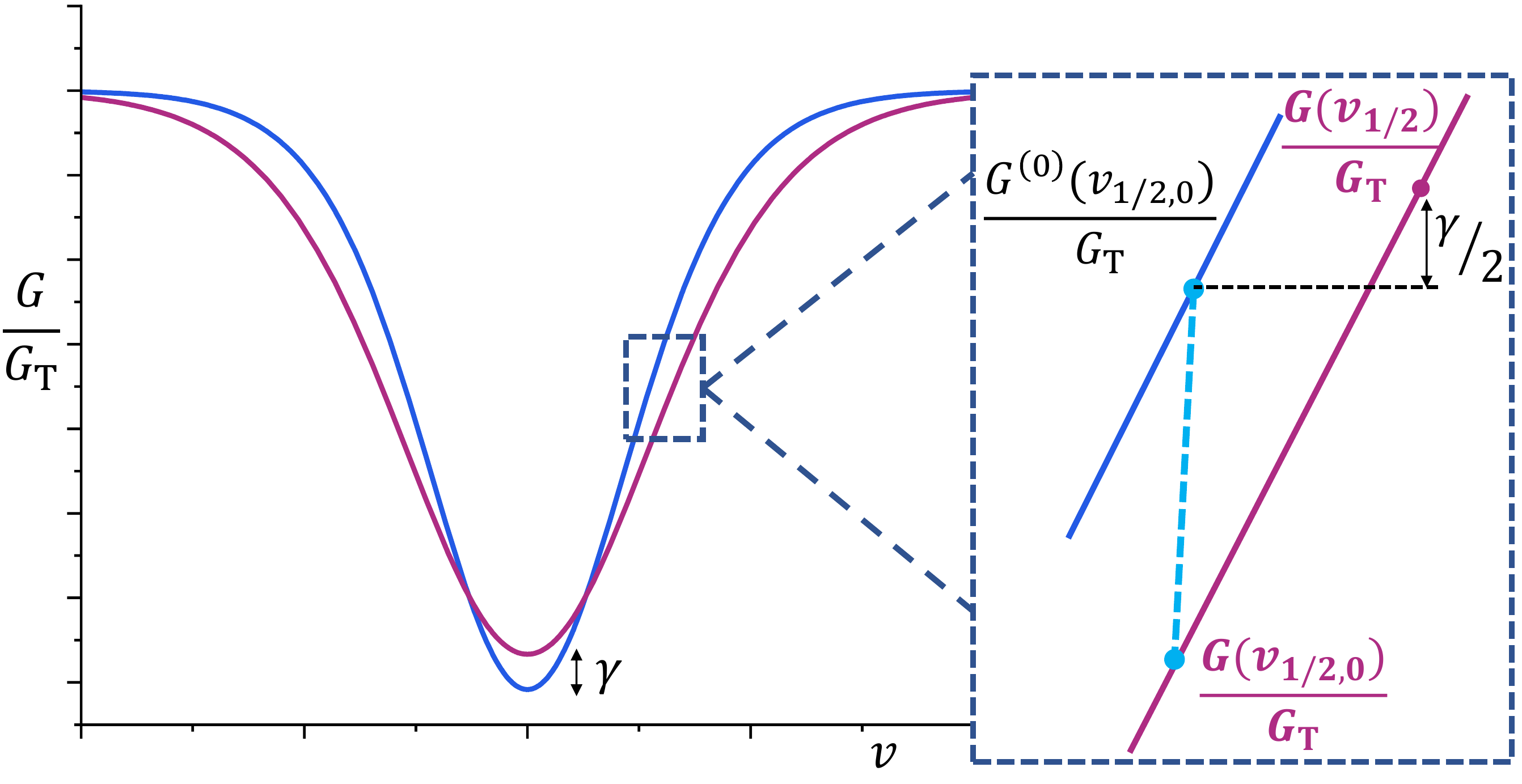}
	\caption{Graphical illustration of how we find correction to half-width for non-uniform arrays and low $T$ corrections. Blue line (deeper and narrower) represents the reference curve and the red one the actual conductance.
		\label{fig2}}
\end{figure}

To obtain the error in temperature in general, we can write the following equations linking the actual halfwidth $v_{1/2}$ to $v_{1/2,0}=2.71959...$, the half-width point of a uniform array from Eq.~\eqref{univ}, i.e. that of the reference curve (see Fig.~\ref{fig2}), as 
\begin{eqnarray}\label{e20}
	&&\frac{G(v_{1/2})}{G_{\rm T}} = \frac{G^{(0)}(v_{1/2,0})}{G_{\rm T}}+\gamma/2\\&& \frac{G(v_{1/2})}{G_{\rm T}}  = \frac{G(v_{1/2,0})}{G_{\rm T}}+ \frac{G'(v_{1/2,0})}{G_{\rm T}} (v_{1/2}-v_{1/2,0}).\nonumber
\end{eqnarray}
Here $\gamma$ is the change of the depth of the conductance curve with respect to the reference one with superscript $(0)$ as shown in Fig.~\ref{fig2}. The error in temperature is then $\delta T/T =\delta v_{1/2}/v_{1/2}$ where $\delta v_{1/2}\equiv v_{1/2}-v_{1/2,0}$.\\
%(i) Resistances vary along the array but capacitances are all equal. This presents the low $T$ sensors, where the island capacitance is much larger than island-to-island (junction) capacitance. (ii) Resistances and capacitances $C_i$ of the junctions vary along the array such that their product is constant. This is the relevant regime for the high $T$ sensors where junction capacitance is the dominant over the island stray capacitance.

\begin{figure*}
	\centering
	\includegraphics [width=0.8\textwidth] {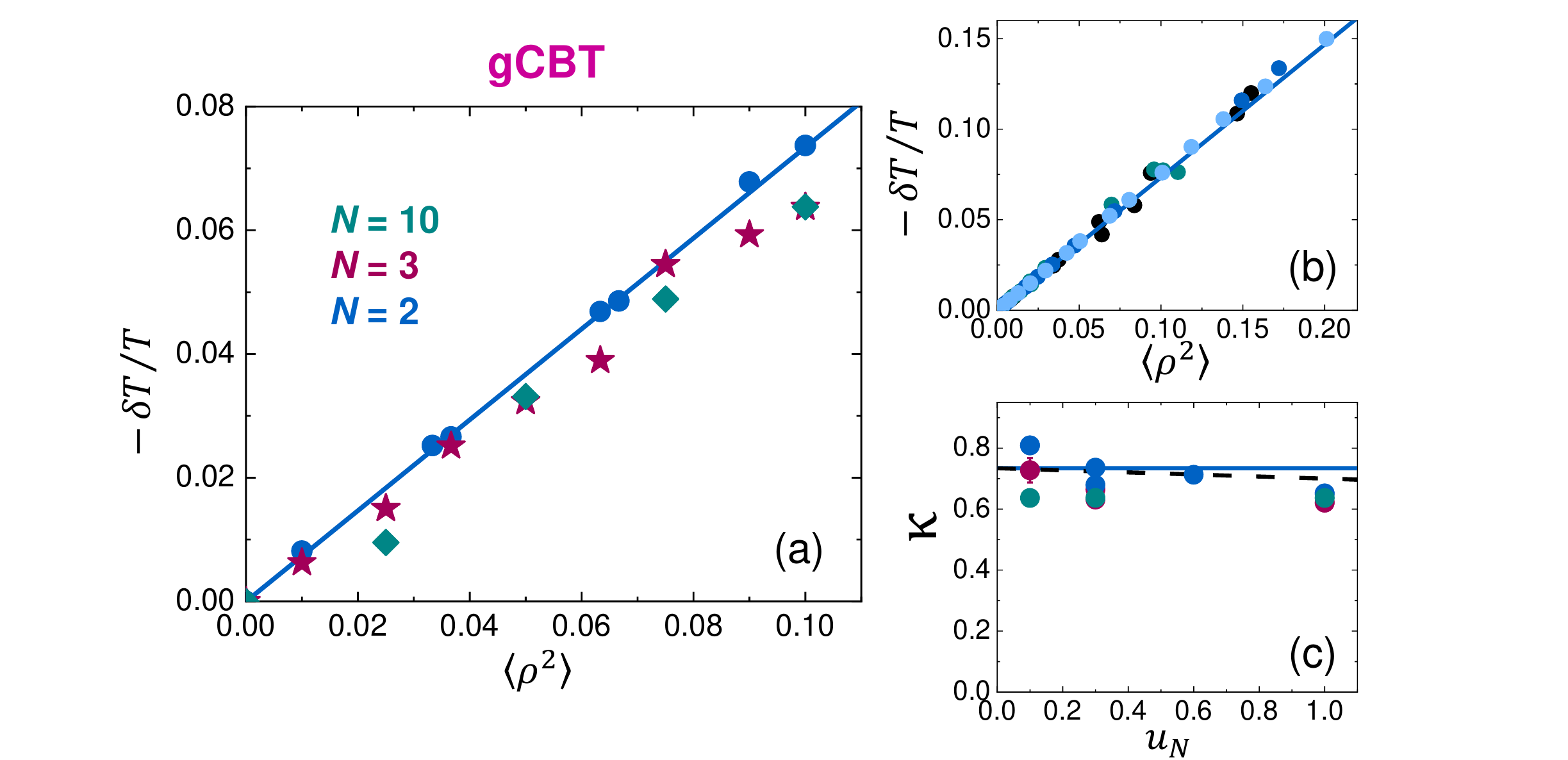}
	\caption{Non-uniformity induced error in temperature reading of gCBT. (a) The relative error $-\delta T/T$ vs. $\langle \rho^2\rangle$ for three different array lengths $N$. The solid line is the analytical result of Eq.~\eqref{e3a} valid for all $N$, and the symbols are calculated numerically with $u_N=0.3$. (b) Numerical results for non-uniformity error in temperature, based on the universal conductance in the high temperature approximation, Eq.~\eqref{univ}. The symbols are for ensembles of random uniform distributions of resistances for: $N=2$ (dark blue), $N=10$ (green), $N=100$ (black), and $N=1000$ (light blue). (c) Here we have $-\delta T/T\equiv \kappa \langle \rho^2\rangle$. The slope $\kappa$ of numerically calculated errors as a function of $u_N$ indicating that the $\langle \rho^2\rangle$ dependent error is close to that in the universal regime (horizontal line) even up to $u_N\sim 1$. The dashed line with a negative slope for $N=2$ is the result of the analytical calculation from Eq. \eqref{e22}. 
		\label{fig2b}}
\end{figure*}

{\bf (i) gCBT:} We find by expanding Eq.~\eqref{univ} for small $\langle \rho^2\rangle$ that
\begin{equation} \label{e2}
G(v)/G_{{\rm T}}=1-u_N g(v)-u_N[g'(v)v+\frac{1}{2}g''(v)v^2]\langle \rho^2\rangle,
\end{equation}
where $u_N\equiv\delta_i/k_{\rm B}T$ is a constant and prime denotes derivative with respect to $v$. Using the argument in Eqs.~\eqref{e20} we then find that for the same value of conductance ($\gamma=0$ here), the bias voltage shifts due to resistance non-uniformity as
\begin{equation} \label{e3}
\delta v_{1/2}= -\big{[}v_{1/2,0}+\frac{1}{2}\,\frac{g''(v_{1/2,0})}{g'(v_{1/2,0})}v_{1/2,0}^2\big{]}\langle \rho^2\rangle.
\end{equation}
The relative error in temperature reading is then
\begin{eqnarray} \label{e3a}
\delta T/T&&\simeq-\big{[}1+\frac{1}{2}\,\frac{g''(v_{1/2,0})}{g'(v_{1/2,0})}v_{1/2,0}\big{]}\langle \rho^2\rangle\nonumber\\&&\simeq -0.734\,\langle \rho^2\rangle,
\end{eqnarray}
where the last form arises from $v_{1/2,0}\simeq 2.7196$ in the universal regime, Eq.~\eqref{univ}. Figure~\ref{fig2b} presents by solid line the analytical result of Eq.~\eqref{e3a} which is valid for all values of $N$.

\begin{figure*}
	\centering
	\includegraphics [width=0.8\textwidth] {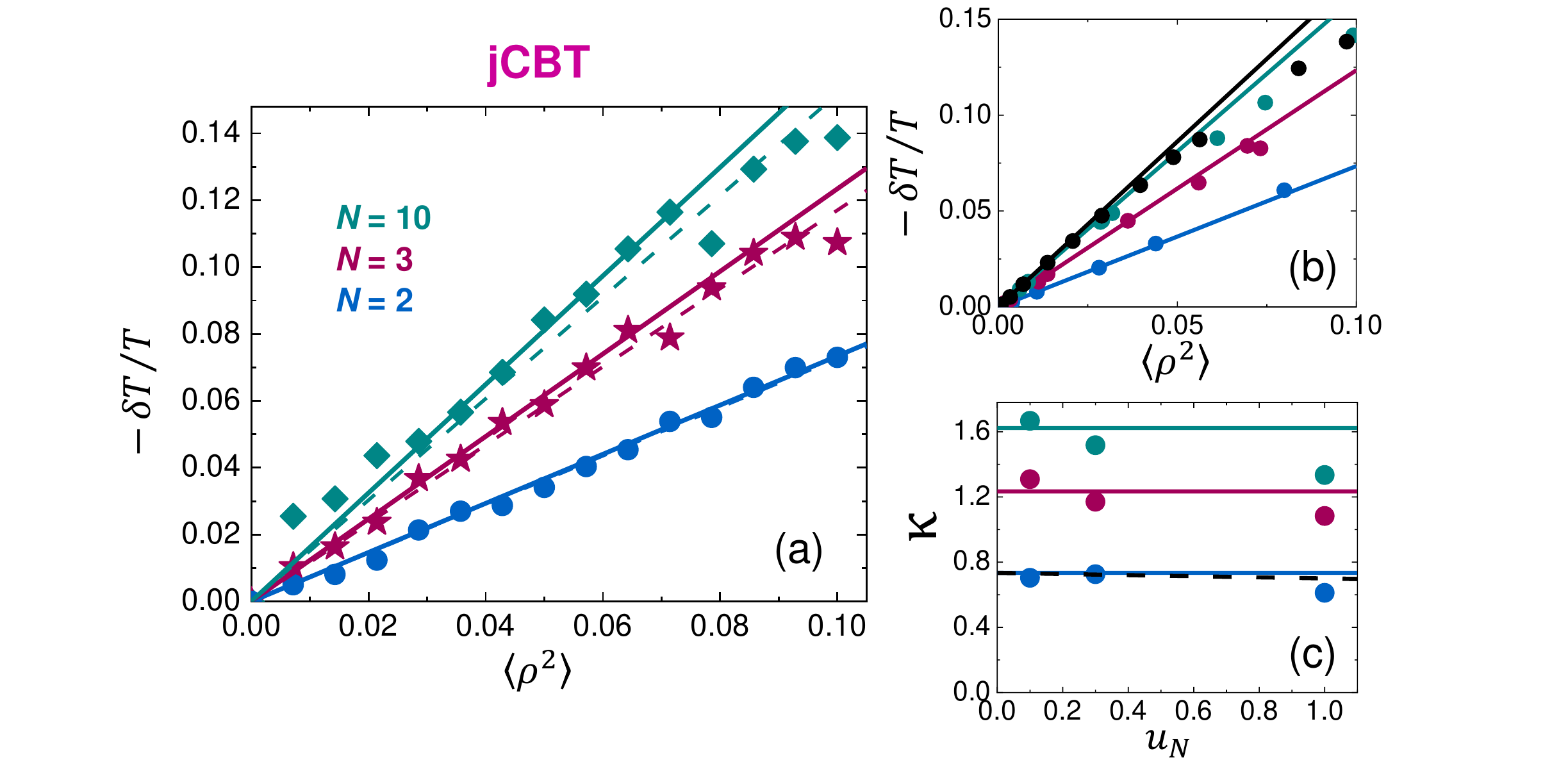}
	\caption{Non-uniformity induced error in temperature reading of jCBT. (a) The relative error $-\delta T/T$ vs. $\langle \rho^2\rangle$ for three different array lengths $N$. The solid lines are the analytical results of Eq. \eqref{e7}, and the symbols are calculated numerically with $u_N=0.3$. The dashed lines are the linear fits to the latter data. (b) Numerical results for non-uniformity error in temperature, based on the universal conductance in the high temperature approximation, Eq.~\eqref{univ}. The symbols are for ensembles of random uniform distributions of resistances for: $N=2$ (blue), $N=3$ (red), $N=10$ (green), and $N=100$ (black). (c) The slope $\kappa$, where $-\delta T/T =\kappa \langle \rho^2\rangle$, of numerically calculated errors as a function of $u_N$ indicating that the $\langle \rho^2\rangle$ dependent error is close to that in the universal regime (horizontal lines, $N=10,3,2$ from top to bottom) even up to $u_N\sim 1$. The dashed line with a small negative slope for $N=2$ is the result of the analytical calculation from Eq. \eqref{e22}. 
		\label{fig3}}
\end{figure*}

{\bf (ii) jCBT:} in this case both the width and the depth of the conductance dip are changing. According to \cite{Hirvi1995,prl2008}, we have for the capacitive term in Eq. \eqref{e1}
\begin{equation} \label{e4a}
\delta_i /e^2=C_{i-1,i-1}^{-1} + C_{i,i}^{-1} - 2C_{i,i-1}^{-1}.
\end{equation}
Ignoring the island capacitance fully, the elements on the right hand side of Eq. \eqref{e4a} read \cite{in1992}
\begin{equation} \label{e4b}
C_{k,l}^{-1} = \tilde C \sum_{m=1}^{{\rm min} (k,l)} \frac{1}{C_m} \sum_{{\rm max} (k,l)+1}^{N} \frac{1}{C_n},
\end{equation}
where $C_i$ is the capacitance of junction $i$ and $\tilde 	C^{-1} =\sum_{k=1}^{N} {C_k}^{-1}$. If we define $C$ such that $R_{{\rm T},i}C_i=R_{\rm ave}C$, we have $\tilde C=C/N$ and  
\begin{equation} \label{e4c}
\frac{\delta_i}{k_{\rm B}T}=\frac{e^2}{k_{\rm B}TC}\big{[}\frac{R_{{\rm T},i}}{R_{\rm ave}}-(\frac{R_{{\rm T},i}}{R_{\rm ave}})^2/N\big{]}.
\end{equation}
With similar approximations as above, we find that the zero bias $v=0$ conductance has the value 
\begin{equation} \label{e5}
{G(0)}/{G_{\rm T}}=1-\frac{e^2}{6k_{\rm B}TC}\frac{N-1}{N}\big{[}1+\frac{N-3}{N-1}\langle \rho^2\rangle\big{]}.
\end{equation}
At finite $v$ we have
\begin{eqnarray} \label{e6}
{G(v)}/&&{G_{\rm T}}=1-\frac{e^2}{k_{\rm B}TC}\bigg{\{}\frac{N-1}{N}g(v)+\big{[}\frac{N-3}{N}g(v)\nonumber\\&&+\frac{2N-3}{N}vg'(v)+\frac{N-1}{2N}v^2g''(v)\big{]}  \langle \rho^2\rangle\bigg{\}}.
\end{eqnarray}
Note that $u_N=\frac{e^2\langle 1/C\rangle}{k_{\rm B}T}\frac{N-1}{N}$ in this case, where $\langle 1/C\rangle$ is the average of inverse junction capacitances $C_i$. Again using Eqs.~\eqref{e20} we find
\begin{eqnarray} \label{e7}
\delta T/T&&\simeq -\big{[}\frac{2N-3}{N-1}+\frac{1}{2}\,\frac{g''(v_{1/2,0})}{g'(v_{1/2,0})}v_{1/2,0}\big{]}\langle \rho^2\rangle\nonumber\\&& \simeq -\big{[}\frac{2N-3}{N-1}-0.265945\big{]}\langle \rho^2\rangle.
\end{eqnarray}
Here the last step arises since $\frac{g''(v_{1/2,0})}{2g'(v_{1/2,0})}v_{1/2,0}\simeq -0.265945$. Unlike for gCBT in (i), here the error depends on $N$. We note that the results of (i) and (ii) are equal for $N=2$ as they should.
Figure \ref{fig3} presents by solid lines the analytical results on non-uniformity error in jCBT for different values of $N$. Equations~\eqref{e3a} and~\eqref{e7} are the main results of the paper in the universal regime, $E_{\rm C}\ll k_{\rm B}T$.

\section{Beyond the linear regime}
To obtain the linear in $u_N$ results above, the actual charge distribution on the islands plays no role. This, however, is not the case at low relative temperatures, $k_{\rm B}T \lesssim E_{\rm C}$. To see how the conductance given by Eq. \eqref{univ} gets modified in this case, in partucular to find the corresponding expression up to $u_N^2$, we take the simple two-junction device. In the following we write $u\equiv u_2$ for brevity. First we find the charge distribution, i.e. the occupation probability $\sigma(n)$ for different electron numbers on the island, which is governed by the solution of the master equation
\begin{eqnarray} \label{e8}
\dot \sigma(n)&&= \big{[}\Gamma_1^+(n-1)+\Gamma_2^-(n-1)\big{]}\sigma(n-1)\\&&+\big{[}\Gamma_1^-(n+1)+\Gamma_2^+(n+1)\sigma(n+1)\big{]}\nonumber\\&&-\big{[}\Gamma_1^+(n)+\Gamma_1^-(n)+\Gamma_2^+(n)+\Gamma_2^-(n)\big{]}\sigma(n)\nonumber
\end{eqnarray} 
in steady state $\dot\sigma(n)=0$. Here $\Gamma_i^\pm (m)$ is the tunneling rate in junction $i=1,2$ in either forward ($+$) or backward ($-$) direction with $m$ extra electrons on the island. The key idea here is to assume that the distribution in the thermometer is broad such that the occupation $\sigma(n)$ is a smooth function of $n$, extending over many possible values of $n$ such that it can be taken as a continuous variable~\cite{jltp1997}. In fact the variance of the electron number (at zero bias voltage) is simply $\langle \delta n^2\rangle = 1/u$ thus becoming very wide for $E_{\rm C}\ll k_{\rm B}T$. Expanding Eq.~\eqref{e8} in $n$ yields
\begin{eqnarray} \label{e9}
&&[(\Gamma_2^+(n)-\Gamma_2^-(n))-(\Gamma_1^+(n)-\Gamma_1^-(n))]\sigma(n)
\\&&+\frac{\partial}{\partial n}\big{\{}[\Gamma_1^+(n)+\Gamma_1^-(n)+\Gamma_2^+(n)+\Gamma_2^-(n)]\sigma(n)\big{\}}=0.\nonumber
\end{eqnarray}
To obtain the rates $\Gamma_i^\pm(n)$ we write the energy cost of each event for the system biased at voltage $V$ as $\delta F_i^\pm=\pm eV_i+\delta E_i^\pm(n)$, where $V_i=(R_i/R_\Sigma) V$ is the voltage drop across each junction with resistance $R_i$, and $R_\Sigma=R_1+R_2$. $\delta E_i^\pm(n)$ is the change of the charging energy $E_{\rm C}=(ne)^2/(2C_\Sigma)$ ignoring the offset charges (validated by broad variation of $n$). Here $C_\Sigma\equiv C_1+C_2$. Normalizing the energies as $\delta f_i^\pm \equiv \delta F_i^\pm /(k_{\rm B}T)$, we can write for each event then $\delta f_1^+(n)=v_1+\delta\epsilon_1^+$, $\delta f_1^-(n)=-v_1+\delta\epsilon_1^-$, $\delta f_2^+(n)=v_2+\delta\epsilon_2^+$, $\delta f_2^-(n)=-v_2+\delta\epsilon_2^-$, where $v_i=eV_i/(k_{\rm B}T)$, $\delta\epsilon_1^+=\delta\epsilon_2^-=(1/2+n)u$ and $\delta\epsilon_1^-=\delta\epsilon_2^+=(1/2-n)u$. The rates themselves are given by the standard expression for normal metal junctions~\cite{in1992} as
\begin{equation} \label{e11}
	\Gamma_i^\pm (n)=\frac{1}{e^2R_i}\frac{\delta F_i^\pm(n)}{1-e^{-\delta F_i^\pm(n)/(k_{\rm B}T)}}.
\end{equation}
Our strategy is to expand the rates in powers of $u$ to obtain the results for the thermometer in its working regime $u\lesssim 1$. In the leading order in $u$ we then obtain
\begin{eqnarray} \label{e12}
	&&[\Gamma_2^+(n)-\Gamma_2^-(n)]-[\Gamma_1^+(n)-\Gamma_1^-(n)] =\frac{k_{\rm B}T}{e^2}\big{[}(\frac{1}{R_1}+\frac{1}{R_2})n \nonumber\\&&-\frac{1}{2}(\frac{1}{R_1}-\frac{1}{R_2})+(\frac{q(v_1)}{R_1}-\frac{q(v_2)}{R_2})\big{]}u,
\end{eqnarray}
where $q(x)=[1-(1+x)e^{-x}]/(1-e^{-x})^2$. Similarly we obtain
\begin{eqnarray} \label{e13}
	&&\Gamma_1^+(n)+\Gamma_1^-(n)+\Gamma_2^+(n)+\Gamma_2^-(n)=\frac{k_{\rm B}T}{e^2}\bigg{[}\frac{h(v_1)}{R_1}+\frac{h(v_2)}{R_2}\nonumber\\&&+\big{\{}2n\big{(}\frac{q(v_2)}{R_2}-\frac{q(v_1)}{R_1}\big{)}+n(\frac{1}{R_1}-\frac{1}{R_2})-\frac{1}{2}(\frac{1}{R_1}+\frac{1}{R_2})\big{\}}u\bigg{]}\nonumber\\,
\end{eqnarray}
where $h(x)\equiv x\coth(x/2)$.
Inserting Eqs. \eqref{e12} and \eqref{e13} into \eqref{e9} we obtain
\begin{equation} \label{e14}
	u(1/R_1+1/R_2)n\sigma(n) +\frac{1}{2}(h(v_1)/R_1+h(v_2)/R_2)\sigma'(n)=0.
\end{equation}
Here we have ignored contributions proportional to $u\delta R$ as small, where $\delta R \equiv(R_1-R_2)/2$. Equation \eqref{e14} yields a Gaussian distribution, which by normalization $\int_{-\infty}^{\infty} \sigma(n)dn=1$ reads
\begin{equation} \label{e15}
	\sigma(n) =\sqrt{\frac{(\frac{1}{R_1}+\frac{1}{R_2})u}{\pi (\frac{h(v_1)}{R_1}+\frac{h(v_2)}{R_2})}}\exp\bigg{(}-\frac{(\frac{1}{R_1}+\frac{1}{R_2})u}{\frac{h(v_1)}{R_1}+\frac{h(v_2)}{R_2}}n^2\bigg{)}.
\end{equation}
The procedure to obtain the conductance $G$ of the thermometer is to write the current through each junction as 
\begin{equation} \label{e16}
	I_i(v_i)= e\int [\Gamma_i^+(n)-\Gamma_i^-(n)]\sigma(n) dn,
\end{equation}
The conductance of junction $i$, $G_i$, yields the conductance of the thermometer as $G=dI/dV=G_1G_2/(G_1+G_2)$. Taking terms up to $u^2$ we find for junction 1
\begin{eqnarray} \label{e18}
	G_1=\frac{R_2}{R_1+R_2}\big{\{}&&1-ug(v_1)-\frac{u^2}{4}\big{[}g''(v_1)(\frac{h(v_1)}{R_1}+\frac{h(v_2)}{R_2})\nonumber\\&&+g'(v_1)(\frac{h'(v_1)+h'(v_2)}{R_1})\big{]}\big{\}},
\end{eqnarray} 
and $G_2$ can be obtained by permuting the indices 1 and 2. We then finally have up to $u^2$
\begin{widetext}
\begin{eqnarray} \label{e19}
	G(v)/G_{\rm T}= &&1-u g(v)-\frac{u^2}{4}[g'(v)h'(v)+g''(v)h(v)]\nonumber\\&&-\bigg{\{}u[v g'(v)+\frac{1}{2}v^2g''(v)]+\frac{u^2}{4}[4v^2g'(v)^2-g'(v)h'(v)+vg'''(v)h(v)-vg''(v)h'(v)\nonumber\\&&+\frac{1}{2}v^2g''''(v)h(v)+\frac{1}{2}v^2g'''(v)h'(v)+\frac{1}{2}v^2g''(v)h''(v)+\frac{1}{2}v^2g'(v)h'''(v)]\bigg{\}}\,\langle\rho^2\rangle.
\end{eqnarray}
\end{widetext}
In the next sections we use the result of Eq. \eqref{e19} with two different aims: to evaluate the second order corrections to the conductance curve and its width for both uniform~\cite{jltp1997} and non-uniform CBT.
\subsection{Correction to halfwidth in a uniform array due to non-vanishing $u$}
Here we investigate the uniform array within the second order approximation of Eq. \eqref{e19}. Then the lengthy contribution proportional to $\langle\rho^2\rangle$ vanishes, and we have
\begin{equation} \label{e19u}
G(v)/G_{\rm T}= G^{(0)}(v)/G_{\rm T}-\frac{u^2}{4}[g'(v)h'(v)+g''(v)h(v)].
\end{equation}
where $G^{(0)}(v)/G_{\rm T}\equiv 1-u\,g(v)$ yields the standard halfwidth $v_{1/2,0}$ for a vanishingly small $u$. 

The known \cite{jltp1997} lowest order correction to $v_{1/2,0}$ of a uniform CBT due to non-vanishing $u$ can again be obtained with the help of Eqs. \eqref{e20} and Fig. \ref{fig2}. Including the second order correction to $G$ in Eq. \eqref{e19u} suppresses partly the depth of the dip $\Delta G/G_{\rm T}$ from $u/6$ to $u/6-u^2/60$, i.e. $\gamma=u^2/60$. 
Up to linear order in $u$ we can then write the solution for the error in temperature as
\begin{widetext}
\begin{equation} \label{e21} 
\frac{\delta T}{T}=\frac{v_{1/2}}{v_{1/2,0}}-1= -\frac{1+30[g'(v_{1/2,0})h'(v_{1/2,0})+g''(v_{1/2,0})h(v_{1/2,0})]}{20g'(v_{1/2,0})v_{1/2,0}} \frac{\Delta G}{G_{\rm T}} \simeq 0.3921  \frac{\Delta G}{G_{\rm T}},
\end{equation}
\end{widetext}
where $\Delta G/G_{\rm T}$ is the measured depth of the dip.

\subsection{Correction to halfwidth from non-uniformities for non-vanishing $u$}
We consider for simplicity the case $N=2$ and use again Eqs. \eqref{e20} to obtain the error in temperature reading via deformation of the $G/G_{\rm T}$ vs. $v$ analytically. If we write Eq. \eqref{e19} in the form $G/G_{\rm T}=1-ug(v)+u^2 L(v)+[u J(v)+u^2B(v)]\langle \rho^2\rangle$ with obvious notations for $L, J$ and $B$ we find the temperature error arising from the non-uniformity in this case as
\begin{equation} \label{e22}
\delta T/T = -\frac{J(v_{1/2}) +uB(v_{1/2})}{v_{1/2}[g'(v_{1/2})+uL'(v_{1/2})]}\,\langle \rho^2\rangle.
\end{equation}
Up to linear order in $u$, the non-uniformity correction is $\delta T/T=-(0.734-0.0341\,u)\,\langle \rho^2\rangle$.

\section{The full error budget of \lowercase{g}CBT}
Equations \eqref{e21} and \eqref{e22} yield the analytical expression of how the temperature error depends on both $\langle \rho^2\rangle$ and $u$ up to linear order in $\Delta G/G_{\rm T}\simeq u/6$ given by
\begin{equation} \label{e23}
	\frac{\delta T}{T} = 0.3921 \frac{\Delta G}{G_{\rm T}} -(0.734-0.205 \frac{\Delta G}{G_{\rm T}})\,\langle \rho^2\rangle.
\end{equation}
The two contributions have naturally a fully different position as error sources. The first part of this, $0.3921 \Delta G/G_{\rm T}$, given by Eq. \eqref{e21} is a systematic error for both gCBT and jCBT that one can naturally correct for by measuring the $\Delta G/G_{\rm T}$. On the contrary, the second part proportional to $\langle \rho^2\rangle$ remains as uncertainty that is difficult to correct for, and it depends on the fabrication uniformity of the CBT sensors. The coefficients $0.3921$ and $0.734$ in gCBT are valid for $N > 2$ as well, whereas the coefficient $0.205$ was calculated here only for $N=2$ using Eq.~\eqref{e22}.  

It is worth noting that the non-uniformity of gCBT does not lead to corrections in the depth of the zero bias peak. This is seen for instance by setting $v=0$ in Eq.~\eqref{e19}: all the corrections $\propto \langle \rho^2\rangle$ vanish then; this may be a helpful result when using the secondary mode of CBT, i.e. when measuring the depth $\Delta G/G_{\rm T}$, for thermometry.

\section{Numerical Monte Carlo method}

The main features of numerical Monte Carlo simulation are described in \cite{Bakhvalov,Yurttagul2020,Wasshuber}. Solving island potentials and potential differences between islands is done somewhat differently from what is presented in Ref.~\cite{Yurttagul2020}.
We assume that one end of the $N$-junction array is at the ground potential and the other one at potential $V$. Let $\bm{\varphi} = \begin{bmatrix} \varphi_1 & \varphi_2 & \dots & \varphi_{n-1} \end{bmatrix}^T$ denote the potential of each island and $\bm{\tilde{\varphi}} = \begin{bmatrix} \tilde{\varphi}_1 & \tilde{\varphi}_2 & \dots & \tilde{\varphi}_{n} \end{bmatrix}^T$ is the vector of the voltage across each junction. We have then $V=\sum_{i=1}^{n} \tilde{\varphi}_i$ and $\tilde{\varphi}_{i}=\varphi_{i}-\varphi_{i+1}$. Ignoring the offset charges these relations can be expressed as a matrix equation 
\begin{equation}
     \mathbf{C} \cdot \begin{bmatrix} \bm{\tilde{\varphi}} \\ \bm{\varphi} \end{bmatrix} = \begin{bmatrix} V \\ \bm{q} \\ 0 \\ \vdots \\ 0\end{bmatrix},
\end{equation}
where 
\begin{equation}
    \mathbf{C} =  \begin{bmatrix}
        1       & 1       & \dots   & \dots   &   1     &         &         & \\
        C_1     & -C_2    & 0       &         &         &C_{g}&         & \\
                & \ddots  & \ddots  & \ddots  &         &         & \ddots  & \\
                &         & C_{n-1} & -C_{n}  & 0       &         &         & C_{g} \\
                & 1       &         &  0      & -1      & 1       &         & \\
                &         & \ddots  &         &  \ddots &  \ddots & \ddots  & \\
                &         &         & \ddots  &         & \ddots  &  \ddots &  1\\
                &         &         &         & 1       &         &   0     & -1 \\ 
    \end{bmatrix}.
\end{equation}
The vector of island charges is $\bm{q} = \begin{bmatrix} q_1 & q_2 & \dots & q_{n-1} \end{bmatrix}^T$. This procedure allows one to find the island potentials $\varphi_i$. The rest of the simulation is similar to that in~\cite{Yurttagul2020}, determining conductance from current. 

Since we are looking for small deviations in the conductance curves, one needs to average the simulated measurement of current versus voltage over a sufficiently long time. This is particularly important for small values of $E_{\rm C}/k_{\rm B}T$, where the zero-bias drop of conductance is small. Typically this means that one needs to simulate $n = 10^9 – 10^{10}$ tunneling events to obtain sufficiently low statistical error in the data of Figs.~\ref{fig2b} and~\ref{fig3}. If one wants to convert this to what it would mean in real measurement time in experiment, one first observes that in the CBT regime each tunneling occurs in an average time of $\sim e^2R_i/k_{\rm B}T$; therefore the total time that such a simulation corresponds to is $t\sim n e^2R_i/k_{\rm B}T$. One can see that for $R_i \sim 10$ k$\Omega$ and $T\sim 0.1$ K, this would then correspond to seconds of measuring time. Yet even with fast hardware the simulation of such a large number of tunneling events takes tens of hours. Therefore, these calculations are not feasible without sufficient parallelization. In our case the simulations were realized by Aalto University School of Science "Science-IT" computer resources, allowing for approximately 1000 simulations running in parallel. 

Besides the analytical results, Fig.~\ref{fig2b} presents results on Monte-Carlo simulations addressing $\langle \rho^2\rangle$ and $u$ (i.e. $\Delta G/G_{\rm T}$) dependence of this error. In general one can say that the lowest order results presented in Eqs.~\eqref{e3a} and \eqref{e21} are in practise sufficient to address the errors of gCBT up to $u\sim 1$. Similarly, Fig.~\ref{fig3} includes Monte-Carlo results for jCBT with similar conclusions.

\section{Discussion}
The results presented in this paper are useful for assessing errors in Coulomb blockade thermometry both at very low temperatures, down to sub-mK regime as well as at high temperatures approaching the ambient. In the first case, low $T$, we observe that the concept of gCBT works generally, and the uncertainty arrises only from the resistance non-uniformity. Furthermore, since the structures are physically large for low temperature CBTs (lower $E_{\rm C}$), the variance $\langle \rho^2\rangle$ is also quite small due to smaller relative variations in junction sizes. It is in place to observe that 1\% rms-variation in $R_i$ leads to $<10^{-4}$ error only. On the other hand, the sensors in higher temperature range belong rather to jCBT category where the dominant capacitance is that of the junctions. The higher the temperature, the smaller the junctions are in pursuit of maximum $E_{\rm C}$. This is because for practical purposes the depth of the conductance dip $\Delta G/G_{\rm T}\propto E_{\rm C}/k_{\rm B}T$ needs to be of the order of $10^{-2}$ or so, otherwise the signal-to-noise ratio would be compromised. Small average junction size leads to inevitable variation in these sizes and thus to increased $\langle \rho^2\rangle$. Yet for practical purposes it is good to keep in mind that a rms-variation of $\sqrt{\langle \rho^2\rangle}=10\%$ of junctions leads to an error of only less than 2\% for any length of the array. Finally this paper extended the nonuniformity error analysis beyond the universal $E_{\rm C}\ll k_{\rm B}T$ regime. In particular we made the observation that the temperature error does not change significantly when leaving this universal regime; this conclusion was based on both numerical Monte-Carlo simulations for arbitrary arrays and on analytical results for $N=2$. Error analysis of gCBT has not been presented in literature. Similarly, we presented errors beyond the universal regime here for the first time.

\section{Acknowledgement}
We thank Joonas T. Peltonen for useful discussions. This work was supported by Academy of Finland grant 312057 (QTF Centre of Excellence), European Microkelvin Platform (EMP, No. 824109 EU Horizon 2020) and Real-K project (Grant No. 18SIB02). Real-K received funding from the European Metrology Programme for Innovation and Research (EMPIR) co-financed by the Participating States and from the European Union’s Horizon 2020 research and innovation programme. We thank the Russian Science Foundation (Grant No. 20-62-46026). The numerical calculations were performed using the computer infrastructure within the Aalto University School of Science (Science-IT).

\end{document}